\documentclass{article}[16pt]
\usepackage[utf8]{inputenc}
\usepackage{graphicx}
\begin{document}

\begin{center}
{\Large \bf Problems of cosmology on small scales of the Universe\footnote{This review is based on the report presented for the Scientific Session of
the Physical Sciences Division of the Russian Academy of Sciences (PSD
RAS), on March 19, 2025 (see Physics--Uspekhi 69 (3) (2026); Uspekhi
Fizicheskikh Nauk 196 (3) 238 (2025)).}}

\bigskip

\large {I.D. Karachentsev}

\bigskip

Special Astrophysical Observatory, Russian Academy of Sciences,
Nizhnii Arkhyz, Zelenchukskiy region, Karachai-Cherkessian Republic,
Russian Federation

E-mail: idkarach@gmail.com

{\bf Abstract}
\end{center}
 Six challenges for the standard cosmological model $\Lambda$CDM are listed,
which arise when comparing theoretical predictions with observational data on 
scales of $\sim1$~Mpc. Different parameters of luminous and dwarf galaxies in 
the local sphere with a radius of 12~Mpc are presented. The average densities 
of stellar matter and dark matter are reproduced depending on a distance in the 
Local volume. Observational data on distribution of angular momentum of nearby 
galaxies are considered. A comparison of the dark matter mass estimates for systems 
of galaxies based on motions of their internal (virialized) members and neighboring
galaxies is given. The reasons for the low derived value of the dark matter density,
$\Omega_m = 0.08\pm0.02$, in the Local Universe with respect to the  global  value 
$\Omega_m = 0.30\pm0.02$ are discussed.

    
\section{Introduction}
The standard cosmological model ( $\Lambda$CDM), with a relative
density of baryonic matter of $\Omega_b=0.04$, cold dark matter
density of $\Omega_{\rm DM}=0.26$, and a dominant dark energy
component of $\Omega_{\lambda}=0.70$, successfully describes the observed
structure of the Universe on large scales. Results from
numerical (N-body) simulations of the formation and
evolution of large-scale structure convincingly show that it
represents a cosmic ``web'', where galaxies are concentrated in
``walls'', at the intersections of which extended filaments   
 (chains of galaxies) form. In turn, at the intersections of
cosmic filaments, massive knots~--- galaxy clusters ~--- form,
with their populations being replenished by the influx of
new galaxies along the filaments. As galaxies move relative to
each other in dense regions of clusters, their light gas
components are stripped and collectivized, leading to the
formation of a hot gas intracluster medium that emits
strongly in X-rays. The bulk of the Universe is occupied by
cosmic voids framed by walls and filaments. Dwarf galaxies,
rich in gas and actively forming stars, are occasionally found
in these voids.

All these elements of a large-scale structure are presented
in Fig.~1, which shows the distribution of 5000 nearby galaxies
with radial velocities less than 1500 km~s$^{-1}$ across the sky at
equatorial coordinates. At the center of this map is the Virgo
Cluster, whose virial region is indicated by a black circle. The
Virgo Cluster, located at a distance of 16 megaparsecs (Mpc),
is adjacent to cosmic filaments, forming a starfish pattern. A
vast void (the Local Void) is visible on the left side of the
figure, with an angular size of up to one radian. The Local
Void begins at the boundary of the Local group of galaxies
and extends to a distance of approximately 20~Mpc. Galaxies
with active and quenched star formation are color-coded
according to their morphological type on the de Vaucouleurs
numerical scale, shown in the upper left corner of the figure.
The figure clearly demonstrates that gas-rich galaxies with
young stellar populations predominate in low-density
regions, while the Virgo and Fornax Clusters (located in the
lower right part of the figure) are dominated by galaxies with
old stellar populations, embedded in a massive, hot gas
environment. Through X-ray ``lenses'', only the two marked
clusters would be visible on this map against a generally dark
background. The various properties of galaxy clusters and
their role in cosmology are discussed in detail in the review by
Vikhlinin et al. [1].
\begin{figure*}[h]
\includegraphics[height=8cm]{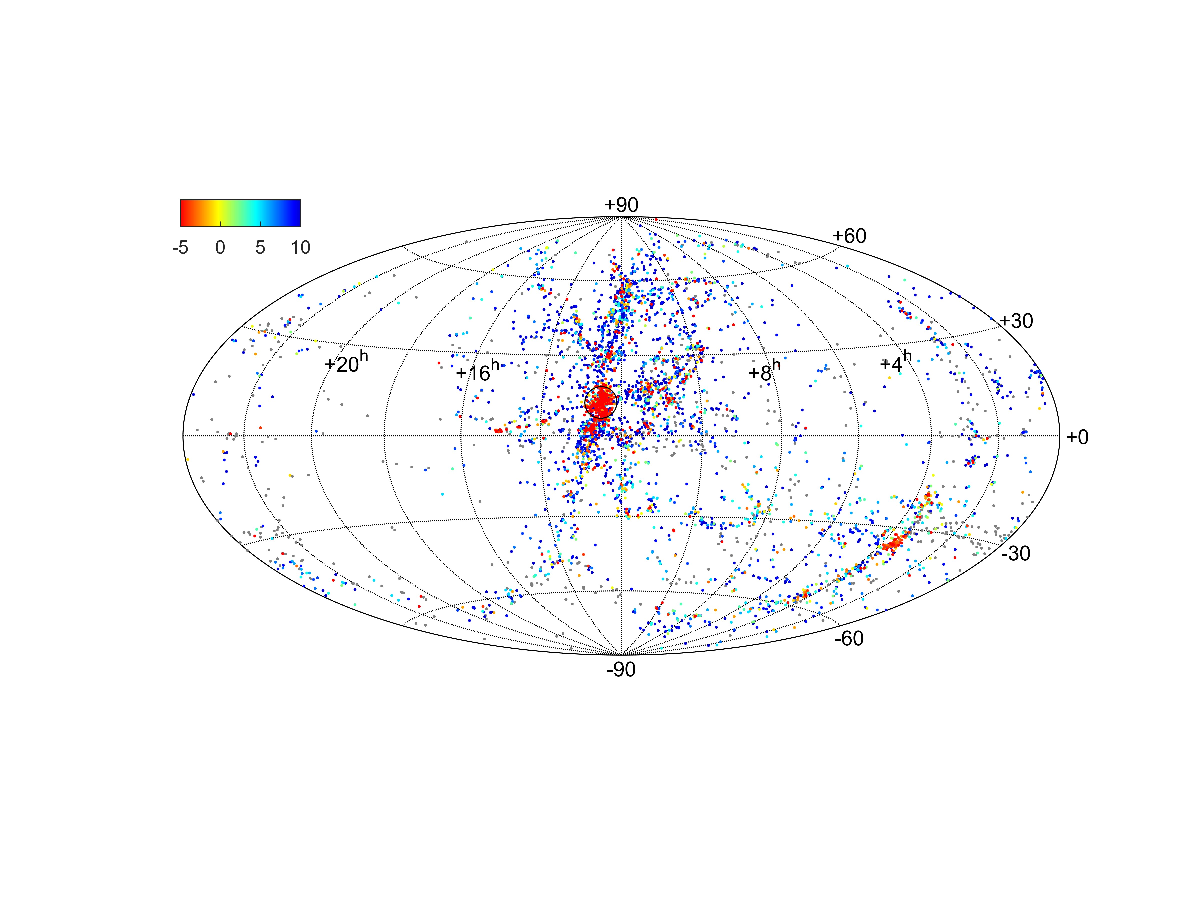}
\caption{Sky distribution of 5000 nearby galaxies with radial velocities less
than 1500 km s$^{-1}$ in equatorial coordinates. Galaxies of different
morphological types on the numerical scale (top left) are shown in
different colors. The virial region of the Virgo cluster is marked by a circle.}
\end{figure*}     

\section{Challenges
for the standard cosmological model}
With increasing computational resources used for N-body
simulations of the dynamical evolution of the Universe, the
spatial resolution of cosmic structures has reached a scale of
$\sim$1 Mpc, or a galaxy mass resolution of about $10^7$ solar
masses [2--4]. However, significant discrepancies have arisen
between simulation results and observational data. Overcoming these discrepancies is an important incentive for
improving the standard cosmological model. Below, we
summarize six main discrepancies discussed in the literature.

\subsection{Lack of dwarf satellites for massive galaxies}
Model calculations yield a rather steep power-law mass
spectrum for dark matter halos in which baryons accumulate
and subsequently form galaxies [5, 6]. The observed number
of small satellites around massive galaxies such as the Milky
Way and the Andromeda Galaxy turned out to be tens of
times smaller than that predicted by $\Lambda$CDM models.
Observational astronomers have conducted systematic
searches for new dwarf satellites using modern deep sky
surveys across various spectral ranges, which has partially
alleviated this discrepancy. On the other hand, theorists have
taken into account that, during their evolution, a significant
fraction of satellites disappeared, being absorbed by larger
galaxies due to dynamical friction [7]. Currently, this
discrepancy persists, although in a less acute form.

\subsection{Excess of thin coplanar structures among satellites}
Over the past decade, numerous publications have appeared
indicating that the satellites of massive galaxies are not
distributed chaotically in space, but instead form thin, planar
structures with coherent satellite motion [8--11]. In particular,
such coplanar subsystems have been discovered in satellites of
the Milky Way, the Andromeda Galaxy (M31), and other
nearby massive galaxies. The characteristic thickness of flat
satellite systems was found to be approximately 20~kpc, with a
total radial extent of 200~kpc. The probability of detecting
such structures based on numerical simulations was found to
be very low. This problem remains unresolved, although
newly discovered satellites are not always embedded in
known planar structures. One possible explanation for the
presence of thin coplanar structures among satellites could be
anisotropic accretion of nearby dwarfs, which are concentrated in cosmic filaments, onto a massive hosts [12].

\subsection{Discrepancy between Hubble parameter estimates}

According to data from the Planck mission, the global value
of the Hubble parameter, which determines the expansion
rate of the Universe and its age, is $H_0=(67\pm1)$~km~s$^{-1}$~Mpc$^{-1}$ 
[13, 14]. In contrast, the local value of the Hubble parameter,
obtained from a comparison of radial velocities and distances
of galaxies, $H_0=(74\pm2)$~km~s$^{-1}$~Mpc$^{-1}$ [15--17], significantly exceeds the global value. This discrepancy between the
$H_0$ estimates is commonly referred to as the ``Hubble tension''.
The local value of the $H_0$ parameter is determined on scales of
$\sim100$~Mpc. When measuring it, it is necessary to carefully
take into account cosmic flows, the amplitudes of which in
different parts of the sky can reach several hundred km~s$^{-1}$.
For example, our Galaxy has an individual velocity of about
630 km~s$^{-1}$ relative to the cosmic microwave background
radiation. Nevertheless, taking into account local flows and
uncertainties at the zero point of the distance scale, the
difference between the local and global values of the $H_0$
parameter significantly exceeds the errors in their measurement. To explain this contradiction, it was suggested that an
observer on the Earth is located in the middle of a huge cosmic
gap measuring $\sim(200-300)$~Mpc [18, 19]. According to
numerical simulations, the voids expand at a rate approximately 10--15\% higher than the global rate. Therefore, an
observer located near the void's center could explain the
higher expansion rate of the local region of the Universe.
However, the standard cosmological model does not predict
average matter density fluctuations with an amplitude greater
than 3\% on scales greater than 200 Mpc

\begin{figure*}[h]
\includegraphics[height=12cm]{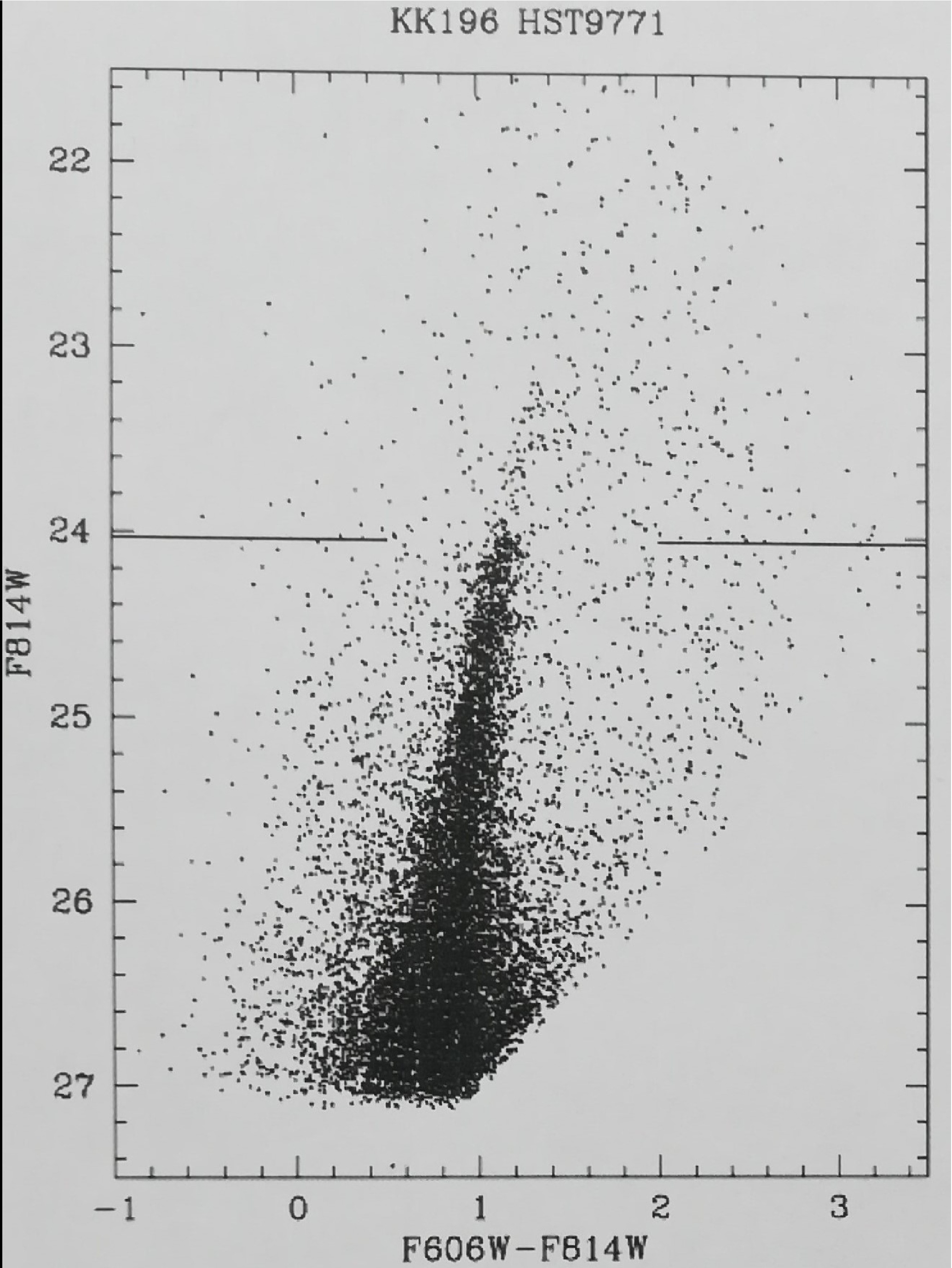}
\caption{Color~---magnitude diagram for stars in the dwarf galaxy KK196
at a distance of 3.96 Mpc, obtained from images in the F814W and F606W
filters taken with the Hubble Space Telescope. The position of the tip of
the red giant branch is indicated by a horizontal line.}
\end{figure*} 

\subsection{Lost baryon problem}
The baryon-to-dark matter ratio  in the $\Lambda$CDM model is  $\Omega_b:\Omega_{\rm DM}\simeq 1:6$.
However, the observed ratio of light (stellar)
to dark matter densities, $\Omega_*:\Omega_{\rm DM}\simeq 1:42$, is approximately
seven times smaller than the expected one [20, 21]. In rich
galaxy clusters, the mass of hot gas exceeds the total mass of
stars by an order of magnitude [1, 22]. Accounting for the
ionized gas of clusters weakens the problem of missing
baryons; however, a discrepancy at the level of a factor of
2--3 still remains. In recent years, a new tool has emerged for
probing the intergalactic medium via dispersion measures and
redshift in fast radio bursts (FRBs). This method allows one
to estimate the average density of free electrons along the
line of sight [23]. Data obtained from studying FRBs have
shown that the intergalactic space is filled with warm ionized
gas, the temperature of which is too low to be detected in the
X-ray range. The average density of this medium turned out
to be quite sufficient to resolve the missing baryon problem
[24, 25].

\subsection{Problem of missing dark matter}
The issue of the existence of dark matter in and between
galaxies has a nearly 90-year history [26, 27]. Observational
manifestations of dark matter are visible in flat (nondecreasing, non-Keplerian) rotation curves at the outskirts of
galaxies, as well as in the large dispersion of radial
velocities in galaxies in groups and clusters. The physical
nature of dark matter remains unknown. A detailed review
of observational data indicating the presence of nonbaryonic dark matter is presented in [27, 28]. The ratio of dark to
light (stellar) matter increases from normal galaxies toward
dwarf galaxies and from binary and triple systems toward
rich groups and clusters. The minimum of this ratio occurs 
in massive galaxies such as the Milky Way. Most estimates of
the dark matter
mass in galaxy groups and clusters are
based on the application of the virial theorem, which
assumes a balance between the kinetic (T) and potential
(U) energies of a system of gravitating bodies: 2T + U = 0.
If the system contains a central dominant galaxy, its total
mass is estimated from the orbital motions of its small
satellites. Recently, new and independent estimates of the
total cluster mass have appeared using the weak gravitational lensing effect, based on an analysis of the pattern of
position angles of distant galaxies around a nearby cluster
with a known redshift. In general, the agreement between
the mass estimates obtained by both methods is quite
satisfactory [29].

Collection of nearby groups and clusters covering the
entire sky and the application of virial mass estimates to them
have shown that the average density of matter contained in
galaxy systems of different scales is $\Omega_m=0.08\pm0.02$ [30--33].
This density is clearly insufficient to reconcile it with the
global value $\Omega_m=0.30\pm0.02$ [13, 14], which follows from an
analysis of the properties of the cosmic microwave background radiation. It is natural to assume that the missing
portion of dark matter is distributed among groups and
clusters of galaxies, but direct methods for its detection have
not yet been proposed. The discussion of this most important
and complex problem will be continued in Section 5.

\subsection{Discrepancy between the expected
and observed orientations of galaxy spins}
The angular momentum of the galaxy rotation, its spin, is a
highly conservative characteristic, remaining constant in
magnitude and direction throughout cosmological time. For
this reason, angular momenta are important indicators of the
initial conditions of galaxy formation. According to theoretical predictions, the spin directions of galaxies are oriented
predominantly in the plane of the wall in which they are
located [34--36]. In particular, galaxies in the Local Supercluster are expected to show alignment of their spins along the
supercluster equator. However, attempts to test this prediction have not yet revealed any significant preference in spin
orientation [37, 38].

\section{Representative sample of galaxies
in the Local volume}
Catalogs of stars and galaxies are usually compiled by
restricting the sample of objects by their apparent magnitude
(flux). The large-scale structure of the Universe is simulated
numerically within a finite volume of space. Therefore, the
results of cosmological simulations are verified using a sample
of galaxies limited by a fixed distance. Compiling such a
sample is associated with great difficulties due to the
enormous range of luminosities of galaxies, as well as the
uncertainties in their distance measurements. For example,
the list of the 1000 brightest galaxies overlaps the list of the
nearest galaxies by less than 10\%. The first list of the Milky
Way's neighbors was compiled in 1979 [39] and included only
179 galaxies with distances less than 10 Mpc of the observer.
Over the years, thanks to targeted efforts [40, 41], the sample
of nearby galaxies has been increased by almost an order of
magnitude. This list continues to be expanded at the present
time, mainly due to the discovery of increasingly faint, low-
surface-brightness galaxies with. The creation of a model and
a representative sample of galaxies in the Local volume
consisted of several main stages.

{\bf 1.}  To search for new nearby low-luminosity galaxies,
large-scale surveys of the northern and southern skies were
used in the optical range (POSS-II, ESO SERC, SDSS, DESI
[42, 43]) and in the 21-cm neutral hydrogen radio line
(HIPASS, Arecibo survey, FAST survey [44--46]).

{\bf 2.} Numerous measurements of the radial velocities of
selected candidate nearby galaxies were carried out using the
6-meter optical telescope (BTA) of the Russian Academy of
Sciences [50], the 100-m radio telescope in Effelsberg,
Germany [47--49], and other major radio telescopes.

{\bf 3.}. The distances to galaxies were determined using the
Hubble Space Telescope (HST) based on the luminosity of the
tip of the red giant branch (TRGB). Using images of the
galaxy in two filters, a color-magnitude diagram for the stars
was constructed. The TRGB position, determined by the
parameters of nuclear reactions in the star interior, made it
possible to measure the distance of any type of galaxy with an
accuracy of $\sim$5\%. As an example, Fig. 2 shows the colormagnitude diagram for the dwarf galaxy KK196, located at a
distance of 3.96 Mpc. In total, high-precision distances for
more than 500 nearby galaxies were measured using images
obtained with the HST.

\begin{table}[b]
\caption{The most massive galaxies of the Local Volume.}
\begin{tabular}{l|r|r|r|r|r}   \hline
Name &   $D$ & $N_v$ & $\sigma_v$                &$M_T$  &            $M_T/M_*$ \\ \hline
    &   Mpc &     & km~s$^{-1}$&  $10^{12}/M_{\odot}$&          \\ \hline
Milky Way&0.01&45&109&1.17&38$\pm$13 \\
M31&0.77&51&113&1.70&48$\pm$9\\
NGC253&3.70&7&42&0.81&13$\pm$4\\
N628=M74& 10.19&9&69&1.44&60$\pm$22\\
NGC891&9.95&5&92&0.83&14$\pm$3\\
NGC1291&9.08&2&121&4.35&75$\pm$7\\
IC342&3.28&8&73&1.59&58$\pm$29\\
NGC2683&9.82&2&43&0.12&3$\pm$3\\
NGC2784& 9.82&2&111&3.04&80$\pm$80\\
NGC2903&9.17&5&41&0.45&11$\pm$9\\
N3031=M81&3.70&31&123&3.10&37$\pm$10\\
NGC3115& 10.20 &10&112&4.82&80$\pm$25\\
NGC3184 &11.12&2&59&0.66&33$\pm$4\\
N3379=M105 &10.80& 33&136&5.75&30$\pm$7\\
NGC3521 &10.70&3&55&0.89&12$\pm$5\\
N3556=M108 &9.90&2&67&1.07&52$\pm$47\\
N3627=M66 &11.10 &22&135&5.89&6$\pm$6\\
NGC4258 &7.66&11&96&2.09 &39$\pm$18\\
Sombrero&9.55&15&96&13.38&91$\pm$35\\
N4736=M94&4.41&15&68&2.40&83$\pm$28\\
N5055=M63&9.04&7&54&0.51&8$\pm$3\\
NGC5128 &3.68&34&124&4.67&58$\pm$17\\
N5194=M51&8.40&5&83&1.35&17$\pm$14\\
NGC5236&4.90&10&61&1.07&24$\pm$6\\
N5457=M101&6.95&8&69&1.07&27$\pm$10\\
NGC6744&9.51&5&71&1.55&24$\pm$14\\
NGC6946&7.73&8&65&1.23&20$\pm$ 8\\ \hline
\end{tabular}
\end{table}  

{\bf 4.}. For Local volume galaxies with distances within 12 Mpc,
the integrated luminosities, stellar and gaseous component
masses, star formation rates, and other characteristics were
determined. A catalog and atlas of $\sim1700$ Local volume
galaxies is presented in the database [51], which is periodically
updated with new objects, while the basic parameters of the
galaxies are continuously refined. As the title page of the database (Fig. 3) shows, the number of visits has already
exceeded 400,000, and the number of citations of the two
published versions of the catalog [40, 41] has reached 1300.
Figure 4 shows the distribution of Local volume galaxies
by stellar mass and distance from the observer. Within a
distance of 12 Mpc, there are 27 galaxies with stellar masses
comparable to the Milky Way (MW) and the Andromeda
Nebula (M31). About half of all fainter galaxies are
associated with these giants. The spread of galaxies in stellar
mass reaches six orders of magnitude. Measurements of the
difference in radial velocities ($\Delta V$) of the companion galaxies
and their projected separations ($R_p$) relative to the main
galaxy allow the total dynamical mass of the group to be determined [52]

\begin{figure*}[h]
\includegraphics[height=10cm]{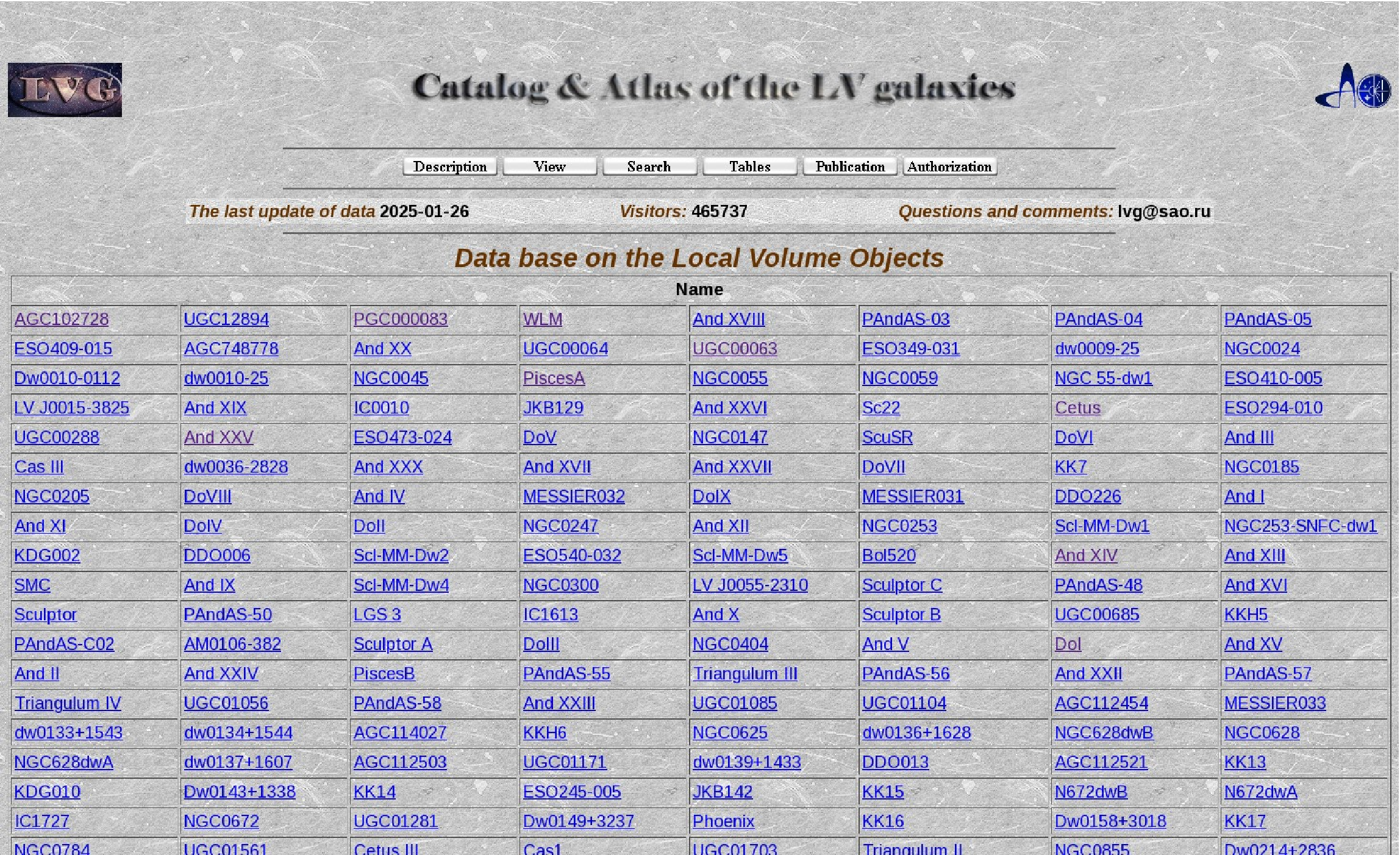}
\caption{Front page of the Local Volume Galaxy Database. As of January 26, 2025, the number of database visits was 465737.}
\end{figure*}

\begin{figure*}[bth]
\includegraphics[height=10cm]{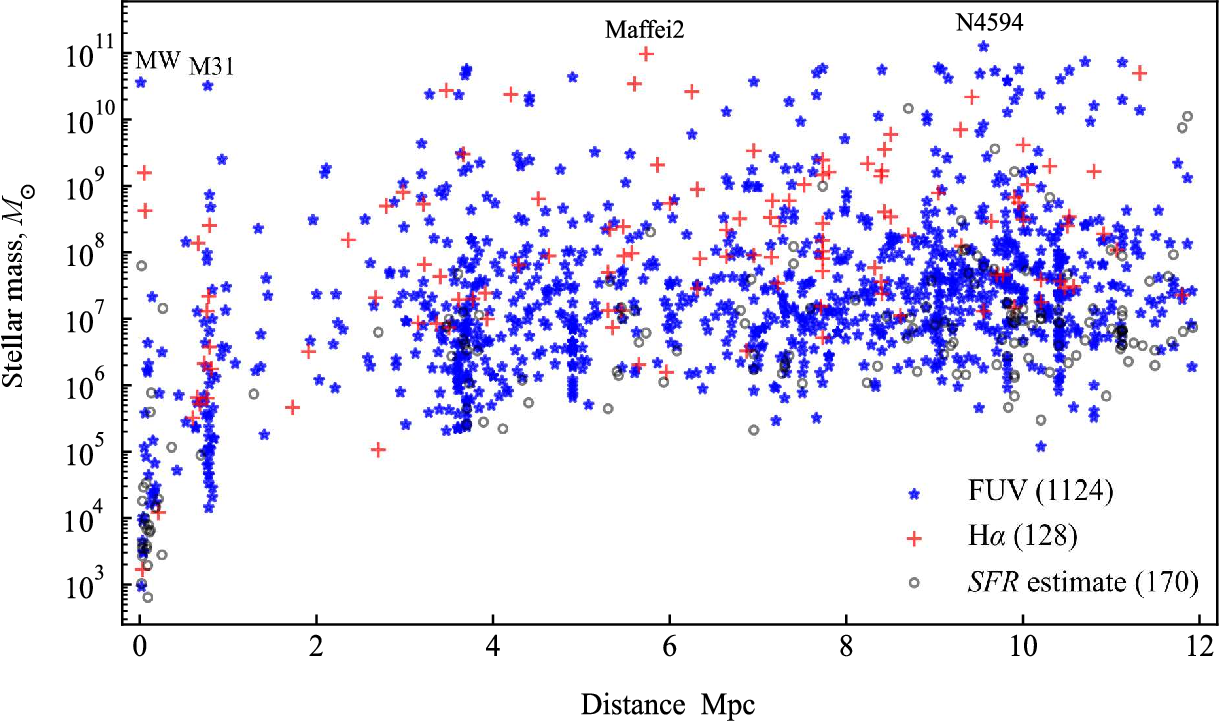}
\caption{Distribution of Local volume galaxies by stellar mass (in solar-mass units) and distance (in Mpc). Different symbols denote galaxies whose star
formation rate determined by far-ultraviolet flux (FUV), H$\alpha$ flux, and the mass-to-SFR calibration relation.}
\end{figure*}

$$M_v=(16\pi/G)\times\langle\Delta V^2\times R_p\rangle, $$
where $G$ is the gravitational constant. Here, the assumption is
made that the orbits of the satellites are oriented randomly
relative to the line of sight, and their average orbital
eccentricity is  $\langle e^2\rangle=1/2$, as follows from the results of
numerical simulations of galaxies in the standard cosmological model.

Table 1 presents a list of the 27 most massive galaxies
in the Local volume. Its columns indicate: (1) the galaxy
name, (2) its distance from the observer in Mpc, (3) the
number of satellites with measured radial velocities, (4) the
dispersion of the radial velocities of the satellites, (5) the
integrated mass of the group, and (6) the ratio of the
integrated mass to the stellar mass of the group. As follows
from these data, typical velocities of galaxy motion in
nearby groups are $\sim100$~km~s$^{-1}$, and the typical ratio of
the total mass to the stellar mass is characterized by a
value of $M_T/M_*\sim40$.

 \section{Orientation of galaxy spins
in the Local volume}
The absolute value of the angular momentum of a galaxy,
defined as the product of the amplitude of its disk rotation by
the characteristic radius of the disk and its baryonic mass,
varies over more than nine orders of magnitude depending on
the luminosity of the galaxy and its morphological type. For a
galaxy arbitrarily oriented relative to the line of sight, the
spatial direction of its spin can be determined by identifying,
from spectral data, which side of its disk is moving away from
the observer. In this case, it is necessary to determine which
side along the minor axis of the galaxy image is closer to the
observer. These conditions can only be met for the nearest
large galaxies, where detailed structure is discernible. In
more distant regions, beyond the Local volume, samples of
galaxies oriented edge-on [37, 53] or face-on, where the
direction of the spiral pattern twist (clockwise or counter-
clockwise) can be identified [54], are used to analyze the spin
orientation. As a result, weak signs of anisotropy in the spin
distribution of spiral galaxies have been detected. However,
questions remained as to whether the observed anisotropy
was related to the location of the galaxies relative to the
nearest walls and filaments of the large-scale structure.
Therefore, we examined the spin orientation of the galaxies
in the Local volume, where the effects of observational
selection and sample incompleteness are minimal, and the
plane of the Local Wall, coinciding with the plane of the
Local Supercluster, is clearly visible.

For each of the 27 major galaxies in the Local volume,
the magnitude and direction of the angular momentum were
determined [38]. The remaining smaller galaxies posess small
angular momenta; the contribution of their absolute angular
momenta to the total sum does not exceed 10\%. The spin
orientation of the 27 major galaxies on the celestial sphere is
shown in Fig. 5 in supergalactic coordinates based on the
data from [38]. The uncertainty in determining the spin
direction is $\sim5^{\circ}$. The size of the circle in the figure is
proportional to the logarithm of the angular momentum.
The distribution of spin directions does not reveal any
preferential alignment toward the plane of the Local Supercluster (SGB$\simeq0$), as predicted by the standard $\Lambda$CDM 
model. Moreover, the average deviation of galaxies from
the plane of the Local Supercluster does not exceed $\pm2$~Mpc
(the SGZ scale in Fig. 5). Thus, we can conclude that, on a
scale of 12 Mpc, the orientation of galaxy spins appears
largely chaotic.

\begin{figure*}[h]
\includegraphics[height=8cm]{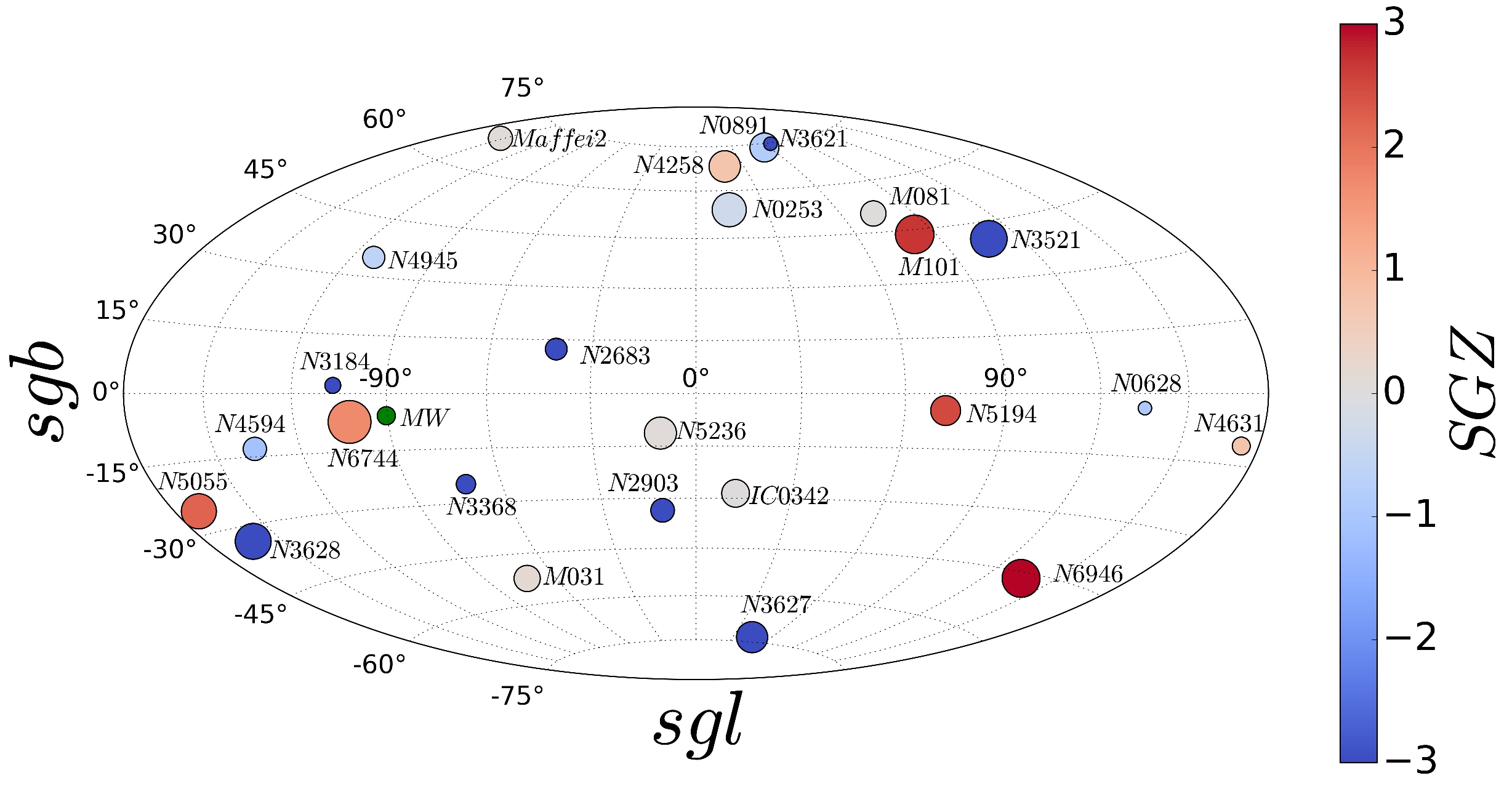}
\caption{Distribution of spin directions of the largest nearby galaxies on
the sky in supergalactic coordinates. The size of the circle is proportional
to the logarithm of the absolute value of the angular momentum. The scale
on the right reflects the galaxy position relative to the plane of the local
``wall'' in Mpc.}
\end{figure*} 
   
   \begin{table*}[h]
\caption{Estimated total mass of galaxy systems.} 
\begin{tabular}{c|c|c|c} \hline
 Systems&$M_v/M_{\odot}$&$M(R_0)/M_{\odot}$ & $R_0$/Mpc\\ \hline
Local group &$(2.8\pm0.6)\times10^{12}$& ($1.9\pm0.2)\times10^{12}$& $0.96\pm0.03$\\
Synthetic group &$(1.8\pm0.5)\times10^{12}$& $(1.7\pm0.2)\times10^{12}$ &$0.93\pm0.03$\\
Virgo Virgo Cluster &$(6.3\pm0.9)\times10^{14}$ &$(7.2\pm0.9)\times10^{14}$ &$7.0\pm0.2$\\
\hline
\end{tabular}
\end{table*}

\section{Local Hubble flow of galaxies,
dark matter, and dark energy}
The total mass of a group or cluster of galaxies can be
determined not only from the internal (virial) motions of
the group/cluster members, but also from the motions of
external (neighboring) galaxies. Figure 6 shows the dis-
tribution of the nearest galaxies by distance from the
dynamic center of the Local Group (located at a distance
of 0.45 Mpc from the Milky Way in the direction of the
Andromeda Galaxy) and their velocity relative to the
center of the Local Group. The vertical cloud on the left
side of the figure is formed by satellites of the Milky Way
and Andromeda with virial velocities of $\sim100$~km~s$^{-1}$.
Neighboring galaxies with distances  $D>1$~Mpc have
positive radial velocities, participating in the general
cosmological expansion. The dotted sloping line in the
figure corresponds to the unperturbed Hubble flow. The
gravitational influence of the Local Group slows down the
motion of neighboring galaxies. The regression line drawn
along them intersects the zero-velocity level at
$R_0\simeq1$~Mpc. This is the radius of the ``zero velocity
sphere'', which separates the collapsing region around the
Local Group from the general cosmological expansion.
The typical ratio of $R_0$ to the virial radius of the group is
$R_0/R_v\simeq3.5$. For a fixed value of the parameter $\Omega_{\lambda}$, the
total mass of the attractor (group or cluster) is related to $R_0$ by the expression

$$M_T(R_0)=(\pi^2/8G)\times H_0^2\times C(\Omega_{\lambda})\times R_0^3,$$

where $C(\Omega_{\lambda}$) is a dimensionless factor ranging from 1 to 9/4.
For $\Omega_{\lambda}=0.7$ and $H_0=74$~km~s$^{-1}$~Mpc$^{-1}$, this expression
takes the form $M_T/M_{\odot}=2.1\times10^{12}\times(R_0/{\rm Mpc})^3$.

\begin{figure*}[h]
\includegraphics[height=12cm]{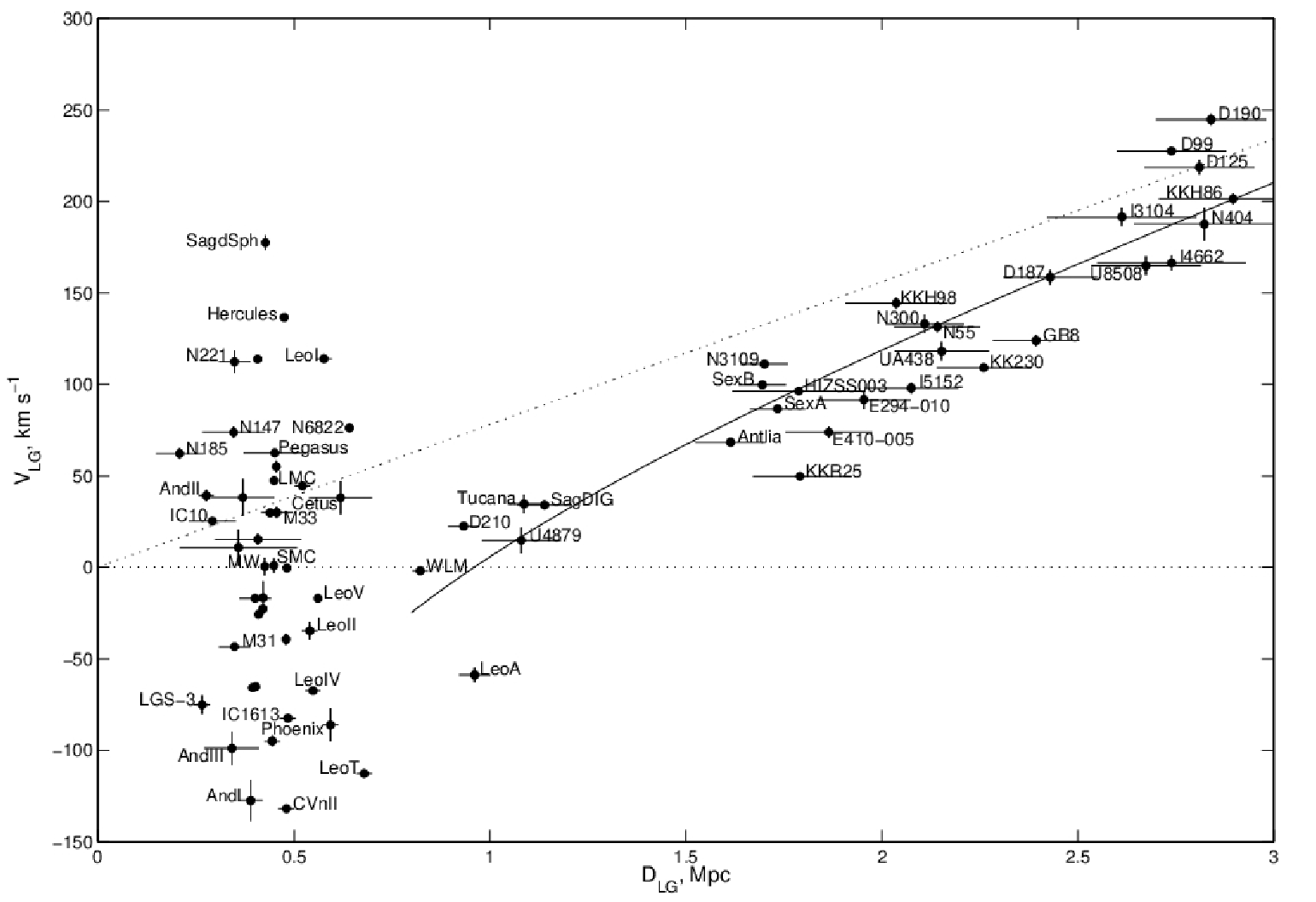}
\caption{Radial velocities and distances of the nearest galaxies relative to the center of mass of the Local Group. Members of the Local Group form a
vertical cloud on the left side of the diagram with virial velocities of approximately 100 km~s$^{-1}$. The slanted dotted line depicts the ideal unperturbed
Hubble flow. Galaxies surrounding the Local Group participate in the general cosmological expansion, but their velocities are slowed by the gravitational
influence of the Local Group.}
\end{figure*}

We have determined the radius  $R_0$ for the Local
Group [55], for a synthetic group composed of several
neighboring groups [56], and for the nearby rich Virgo
cluster [57, 58]. The results are presented in Table 2. As
follows from these data, the estimates of the total mass of
the groups and clusters derived from the virial motions of
the members ($M_v$) agree with the mass estimates from the
motions of neighboring galaxies $M(R_0)$ within the statistical errors of both methods. Note, however, that the
relatively low error in estimating the mass of the Local
Group from the radius $R_0$ is due to the small dispersion
of the radial velocities of galaxies in the neighboring
Hubble flow. According to the latest data [59], the
observed dispersion of the radial velocities of galaxies
falling toward the Local Group is only 15 km~s$^{-1}$, which
is significantly less than the value of 70 km~s$^{-1}$ obtained
from numerical simulations of the local flow. This
discrepancy can be regarded as another small-scale
challenge for the standard cosmological model.

The agreement between the mass values in volumes $(R_v)^3$ and $(3.5R_v)^3$
indicates that the bulk of the dark matter mass is
concentrated within the virial radius of galaxy systems. Note
also that the kinematics of the Hubble flow depends on the
magnitude of the $\lambda$-term, which acts as a universal repulsive
force between galaxies. Furthermore, assuming the absence of
the $\lambda$-term would lead to paradoxical mass estimates within
 $3.5R_v$ that are smaller than those within $R_v$. Here we have
observational evidence of dark energy acting at close distances,
which is important because the evidence for the dark energy
came from observations of extremely distant events~--- supernova explosions that occurred billions of years ago.

Figure 7 shows the distribution of the average stellar mass
density (upper panel) and the total dynamical mass (lower
panel) as a function of the radius of a sphere around the
observer. Data from the UNGC Local volume catalog [41]
and the MK11 galaxy group catalog [32], covering distances
up to 40~Mpc, were used. The horizontal line in the upper
panel corresponds to the global value of the average stellar
matter density derived from deep infrared sky surveys [60]. As
can be seen from these data, on all scales up to 40~Mpc there is
an excess of the local density of luminous matter over its
global value.

\begin{figure*}[h]
\includegraphics[height=12cm]{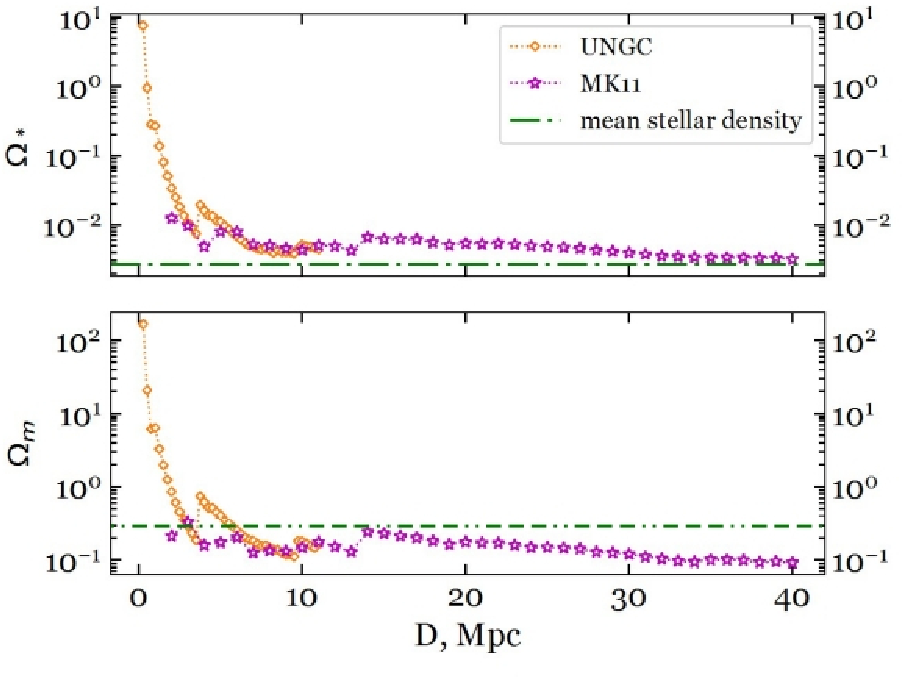}
\caption{Distributions of the mean stellar density (upper panel) and total
matter density (lower panel) as functions of distance from the observer,
based on data from the UNGC [41] and MK11 [32] catalogs. Global values
of the mean stellar density and total matter density are shown as
horizontal lines.}

\end{figure*} 

The behavior of the average density of all matter with
distance $\Omega_m(D)$ generally follows the distribution of $\Omega_*(D)$.
 However, in larger volumes with $D > 20$~Mpc, the average
density  $\Omega_m$ is systematically lower than the global value
 $\Omega_m=0.30$, indicated by the horizontal line. Independent
estimates of the virial masses of galaxy groups and clusters,
based on data from the catalog [33], also yield an average
value of  $\Omega_m\simeq0.08\pm0.02$ on a scale of $\sim40$~Mpc. Various
assumptions have been proposed to explain the discrepancy
between the local and global estimates of $\Omega_m$.

According to [61, 62], the dark matter density profile in
galaxy systems may be more extended than that of baryonic
matter; in this case, most of the dark matter would be
contained in invisible halos surrounding groups and clusters. However, this assumption is contradicted by the
observational data from Table 2, which show an approximate equality between the total mass estimates made within
volumes with radii $R_v$ and $R_0\simeq 3.5 R_v$.

 Another explanation suggests that the 80~Mpc diameter
region of the Local Universe under consideration does not
correspond to the scale of the cosmological ``homogeneity
cell''. An observer on Earth could be located within an
extended cosmic void, where the average matter density is
approximately three times lower than the global density.
However, the data shown in the upper panel of Fig. 7 suggest
that the broader vicinity of our Galaxy is characterized not by
a deficit, but by an excess of local stellar density. Furthermore, galaxy counts extending to deep limits and covering
different regions of the sky [63--66] do not provide any
convincing evidence for the existence of a vast local void
measuring $\sim(100-300)$~Mpc.

A more plausible suggestion is that the bulk of dark
matter is distributed in space between groups and clusters,
either as a homogeneous dark `ocean' or as a population of
dark attractors with varying masses. The dark ocean
hypothesis appears less attractive, since it requires a different
type of dark matter, one not subject to Jeans instability.

The existence of dark attractors with masses of $\sim10^{12}-10^{15}M_{\odot}$,
in which star formation has not occurred,
appears more plausible. Weak gravitational lensing can be
used to detect them. Analysis of the shapes of distant
background galaxies, stretched tangentially relative to a
closer attractor (gravilens), makes it possible to determine
the coordinates and mass of the attractor. A study of this
effect in a 72-square-degree region of the sky, imaged with a
wide-field telescope using subsecond seeing, showed [67] that
among the three hundred detected gravitational potential
peaks, approximately 60\% are not associated with visible
galaxy clusters. This indicates the possible existence of a
population of dark attractors with masses typical of rich
galaxy clusters. Weak gravitational lensing is currently being
observed in deeper surveys covering a significant portion of
the entire sky [14]. Analysis of these data will allow one to
determine in the next few years whether isolated massive dark
matter clumps exist in the Universe.

Note that a diffuse cloud of galaxies, ``Coma I'', has been
discovered in the Local volume, where unusually large
motions of dwarf galaxies are observed, reaching 800 km~s$^{-1}$.
The perturbed kinematics of this region may be due to the
presence of a dark attractor with a mass of  $\sim2\times10^{14}M_{\odot}$,
located at a distance of $\sim$15~Mpc from us [68].

The large-scale structure of the Universe can be roughly
divided into three dynamic categories:

(a) virial zones of groups and clusters, where a balance
between kinetic and potential energies has been established,
and the members of the system have ``forgotten'' the initial
conditions of their formation;

(b) collapsing regions around virial zones, bounded by
zero-velocity spheres with radius $R_0$; and

(c) the remaining infinitely expanding volume of the
general metagalactic field.

The results of numerical simulations of the large-scale
structure [69, 70], together with accumulated observational
data in the Local volume [71], make it possible to characterize
these regions by the parameters presented in Table 3. As can
be seen from the first two rows of the table, approximately
half of all galaxies and more than 80\% of their stellar mass are
contained within the virial zones of groups and clusters.
Apparently, the main stage of the dynamic evolution of the
large-scale structure has already been completed. Moreover,
at the present epoch, the virial volumes of groups and clusters
occupy only 0.1\% of the total volume, and only about 10\% of
the stellar mass is concentrated within the remaining 95\% of
the expanding volume. From these figures, which still retain
significant uncertainties, it can be estimated that the ratio of
dark and stellar matter in the general field outside the virial and
collapsing zones around galaxy systems reaches a value of $M_{\rm DM}/M_*\sim1000$.

\begin{table*}[b]
\caption{Dynamic regions of the cosmic web.}
\begin{tabular}{l|r|r|r}  \hline

 Parameters& Virialized & Collapsing &Expanding\\ \hline
 Proportion of galaxies&54\% &20\% &26\%\\
 Stellar mass&82\%  &8\% & 10\%\\
 Relative volume &0.1\%  &5\%  &95\%\\
 Contribution to $\Omega_m$  &0.06 & 0.02 & 0.22\\ \hline
\end{tabular}
\end{table*}
 
 \section{Conclusions}
 This brief review outlines the main challenges to the standard
cosmological model ($\Lambda$CDM), revealed through comparisons
between its predictions and observational data obtained in
the Local volume of the Universe. Some of these challenges
are gradually losing their significance, while others continue
to stimulate further development of the theory. In recent
years, deep and extensive sky surveys have been planned using
large ground-based and orbital telescopes operating across
various spectral ranges: the Euclid mission [72], the Roman
Space Telescope [73], and the Vera Rubin Observatory [74].
The implementation of these projects will significantly
advance our understanding of the nature of the mysterious
components of the Universe~--- dark matter and dark
energy. In particular, we will be able to determine whether
purely dark massive attractors exist in the space between
galaxy clusters.

What will the Universe look like in hundreds of billions of
years? The answer depends on the still poorly understood
properties of dark matter and dark energy. Numerous studies
have already been devoted to dark energy. Different interpretations of its nature are presented in reviews [75, 76].
Progress in understanding this fundamental physical substance, permeating the entire space of the visible Universe, has
not yet led to a definite conclusion. It should be noted that the
classical Friedmann equation without the $\lambda$-term and with $\Omega_m=1$ 
describes an oscillatory cosmological model, in which
cycles of expansion and contraction of the Universe can
alternate indefinitely. The inclusion of the $\lambda$-term radically
changes the situation. Acting as a universal repulsive force,
dark energy causes galaxies to recede from one another, leads
to an irreversible decrease in the average density of baryonic
matter, increases entropy, and results in the structural
degradation of the Universe. However, this scenario may
prove to be oversimplified if the space containing black holes
possesses a complex topology and if new, unexpected properties of dark energy are discovered in the future.

  \bigskip
  {\bf Acknowledgments.}
The author thanks D.I. Makarov,
V.E. Karachentseva, E.I. Kaisina, O.G. Kashibadze, and
S.S. Kaisin for their collaboration. This work was supported
by the Russian Science Foundation (grant no. 24-12-00277).
 
\bigskip

{\bf References}

1. Vikhlinin A. A. et al. Phys. Usp. 57 317 (2014); Usp. Fiz. Nauk 184 339
(2014)

2. Gross M. A. K. et al. Mon. Not. R. Astron. Soc. 301 81 (1998)

3. Klypin A. et al. Astrophys. J. 516 530 (1999)

4. Knollmann S. R., Knebe A. Astrophys. J. Suppl. 182 608 (2009)

5. Moore B. et al. Astrophys. J. 524 L19 (1999)

6. Klypin A. et al. Astrophys. J. 522 82 (1999)

7. Kim S. Y., Peter A. H. G., Hargis J. R. Phys. Rev. Lett. 121 211302
(2018)

8. Kroupa P., Theis C., Boily C. M. Astron. Astrophys. 431 517 (2005)

9. Ibata R. A. et al. Nature 493 62 (2013)

10. Pawlowski M. S., Kroupa P., Jerjen H. Mon. Not. R. Astron. Soc. 435
1928 (2013)

11. Mart\'{i}nez-Delgado D. et al. Astron. Astrophys. 652 A48 (2021)

12. Dupuy A. et al. Mon. Not. R. Astron. Soc. 516 4576 (2022)

13. Aghanim N. et al. (Planck Collab.) ``Planck 2018 results. VI.
Cosmological parameters'' Astron. Astrophys. 641 A6 (2020)

14. Anbajagane D. et al. Open J. Astrophys. 8 46161 (2025)
DOI:10.33232/001c.146161; arXiv:2502.17677

15. Kourkchi E. et al. Astrophys. J. 902 145 (2020)

16. Riess A. G. et al. Astrophys. J. Lett. 934 L7 (2022)

17. Riess A. G. et al. Astron. J. 861 126 (2018)

18. Keenan R. C., Barger A. J., Cowie L. L. Astrophys. J. 775 62 (2013)

19. B\"{o}hringer H., Chon G., Collins C. A. Astron.Astrophys. 633 A19 (2020)

20. Fukugita M., Peebles P. J. E. Astrophys. J. 616 643 (2004)

21. Spergel D. N. et al. Astrophys. J. Suppl. 170 377 (2007)

22. Bahcall N. A. et al. Astrophys. J. 541 1 (2000)

23. Popov S. B., Postnov K. A., Pshirkov M. S. Phys. Usp. 61 965 (2018);
Usp. Fiz. Nauk 188 1063 (2018)

24. Macquart J.-P. et al. Nature 581 391 (2020)

25. Yang K. B., Wu Q., Wang F. Y. Astrophys. J. Lett. 940 L29 (2022)

26. Zwicky F. Astrophys. J. 86 217 (1937)

27. Mandelbaum R. et al. Mon. Not. R. Astron. Soc. 368 715 (2006)

28. Zasov A. V. et al. Phys. Usp. 60 3 (2017); Usp. Fiz. Nauk 187 3 (2017)

29. van Uitert E. et al. Astron. Astrophys. 534 A14 (2011)

30. Vennik J. Tartu Astron. Obs. Publ. (73) 1 (1984)

31. Tully R. B. Astrophys. J. 321 280 (1987)

32. Makarov D., Karachentsev I. Mon. Not. R. Astron. Soc. 412 2498
(2011)

33. Kourkchi E., Tully R. B. Astrophys. J. 843 16 (2017)

34. Peebles P. J. E. Astrophys. J. 155 393 (1969)

35. Doroshkevich A. G. Astrophysics 6 320 (1970); Astrofizika 6 581 (1970)

36. White S. D. M. Astrophys. J. 286 38 (1984)

37. Antipova A. V., Makarov D. I., Bizyaev D. V. Astrophys. Bull. 76 248
(2021); Astrofiz. Byull. 76 306 (2021)

38. Karachentsev I. D., Zozulia V. D. Mon. Not. R. Astron. Soc. 522 4740
(2023)

39. Kraan-Korteweg R. C., Tammann G. A. Astron. Nachr. 300 181 (1979)

40. Karachentsev I. D. et al. Astron. J. 12 2031 (2004)

41. Karachentsev I. D., Makarov D. I., Kaisina E. I. Astron. J. 145 101 (2013)

42. Abazajian K. N. et al. Astrophys. J. Suppl. 182 543 (2009)

43. Dey A. et al. Astron. J. 157 168 (2019)
 
44. Donley J. L. et al. Astron. J. 129 220 (2005)

45. Haynes M. P. et al. Astron. J. 142 170 (2011)

46. Zhang C.-P. et al. Sci. China Phys. Mech. Astron. 67 219511 (2024)

47. Huchtmeier W. K. et al. Astron. Astrophys. Suppl. Ser. 141 469 (2000)

48. Huchtmeier W. K., Karachentsev I. D., Karachentseva V. E. Astron.
Astrophys. 377 801 (2001)

49. Huchtmeier W. K., Karachentsev I. D., Karachentseva V. E. Astron.
Astrophys. 401 483 (2003)

50. Karachentsev I. D., Chazov M. I., Kaisin S. S. Mon. Not. R. Astron. Soc.
537 L21 (2025)

51. Kaisina E. I. et al. Astrophys. Bull. 67 115 (2012); Astrofiz. Byull. 67
(1) 120 (2012)

52. Karachentsev I. D., Kudrya Yu. N. Astron. J. 148 50 (2014)

53. Dolgosheeva P., Makarov D., Libeskind N. Astron. Astrophys. 698 L8
(2025)March 2026

54. Shamir L. Mon. Not. R. Astron. Soc. 516 2281 (2022)

55. Karachentsev I. D. et al. Mon. Not. R. Astron. Soc. 393 1265 (2009)

56. Kashibadze O. G., Karachentsev I. D. Astron. Astrophys. 609 A11
(2018)

57. Karachentsev I. D. et al. Astrophys. J. 782 4 (2014)

58. Karachentsev I. D. et al. Astrophys. J. 858 62 (2018)

59. Makarov D. et al. Astron. Astrophys. 698 A178 (2025)

60. Driver S. P. et al. Mon. Not. R. Astron. Soc. 427 3244 (2012)

61. Rines K., Diaferio A. Astron. J. 132 1275 (2006)

62. Masaki S., Fukugita M., Yoshida N. Astrophys. J. 746 38 (2012)

63. Djorgovski S. et al. Astrophys. J. Lett. 438 L13 (1995)

64. Bershady M. A., Lowenthal J. D., Koo D. C. Astrophys. J. 505 50 (1998)

65. Totani T. et al. Astrophys. J. 559 592 (2001)

66. Huang J.-S. et al. Astron. Astrophys. 368 787 (2001)

67. Shan H. et al. Astrophys. J. 748 56 (2012)

68. Karachentsev I. D., Nasonova O. G., Courtois H. M. Astrophys. J. 743
123 (2011)

69. Cautun M. et al. Mon. Not. R. Astron. Soc. 441 2923 (2014)

70. Nuza S. E. et al. Mon. Not. R. Astron. Soc. 445 988 (2014)

71. Karachentsev I. D. Astrophys. Bull. 67 123 (2012); Astrofiz. Byull. 67
(2) 129 (2012)

72. Enia A. et al. (Euclid Collab.) Astron. Astrophys. 691 A175 (2024)

73. Wang K. et al. Mon. Not. R. Astron. Soc. 546 staf2253 (2026)
DOI:10.1093/mnras/staf2253; arXiv:2501.16139

74. Rubin D. et al., arXiv:2506.04327; Astrophys. J. , submitted

75. Chernin A. D. Phys. Usp. 51 253 (2008); Usp. Fiz. Nauk 178 267
(2008)

76. Lukash V. N., Rubakov V. A. Phys. Usp. 51 283 (2008); Usp. Fiz. Nauk
178 301 (2008)

 \end{document}